# Two Theorems on Flat Space-Time Gravitational Theories

MARIO CASTAGNINO

*Departamento de Matemática, Facultad de Ciencias Exactas y Naturales,
Universidad Nacional de Buenos Aires, Pabellón 1, Ciudad Universitaria,
Buenos Aires, Argentina*

and

LUIS CHIMENTO

*Departamento de Física, Facultad de Ciencias Exactas y Naturales,
Universidad Nacional de Buenos Aires, Argentina*

### Abstract

The first theorem states that all flat space-time gravitational theories must have a Lagrangian with a first term that is an homogeneous (degree-1) function of the 4-velocity $u^i$, plus a functional of $\eta_{ij} u^i u^j$. The second theorem states that all gravitational theories that satisfy the strong equivalence principle have a Lagrangian with a first term $g_{ij}(x) u^i u^j$ plus an irrelevant term. In both cases the theories must issue from a unique variational principle. Therefore, under this condition it is impossible to find a flat space-time theory that satisfies the strong equivalence principle.

## §(1): Introduction

Too many times we have heard and read that Einstein's general theory of relativity is an exceptional theory in the realm of physics. In fact, gravitation is the only field that requires a curved space-time background to be explained. Nevertheless all formulations of gravitation in flat space-time have had a short life, and never has a flat space theory made a serious challenge to general relativity. Why?

In this paper we try to answer this question. In our opinion it is almost impossible to formulate a macroscopic self-consistent gravitational theory in flat space-time. More precisely: we shall prove that it is impossible to find a covariant theory drawn from a unique variational principle (and therefore self-consistent as all theories with only one axiom), that satisfies the strong equivalence principle (which we consider a soundly based macroscopic principle) in flat space-time.

Therefore, at least in the macroscopic domain, we must reject all heretical theories and accept the true faith that teaches us that space-time is curved, and that the curvature of space-time produces a phenomenon known as the gravitational field.

### §(2): *Variational Principle and Consequences*

We shall try to study the set of all gravitational theories based on a unique hypothesis, i.e., a unique variational principle. We do so because in almost all flat space-time theories additional hypotheses must be stated in order to comply with the three classical experiments (as in [1]) or to eliminate some internal contradictions: so as to make motion and field equations compatible (as in [2]), or motion equations compatible with flat space-time (as in [3]), or to fix a particular gauge where field equations are only valid (as in [4]). On the contrary if we begin with just one variational principle we are sure that we are free of internal inconsistencies; and that is precisely the case of electromagnetism, which is also the best and only model to build a gravitational macroscopic theory.

The electromagnetic action is

$$S_E = \sum \int m \, \eta_{ij} u^i u^j ds + \sum \frac{e}{c} \int A_i u^i ds + \int \mathcal{L}^{(F)} d^4 x \qquad (2.1)$$

where $\eta_{ij}$ is the Lorentz metric tensor, $ds = (\eta_{ij} dx^i dx^j)^{1/2}$, $u^i = dx^i/ds$, etc. The first term in the right-hand side is the free-particle action (i.e., a function of the particle velocity only), the second one is the interaction term (i.e., a functional of the velocity and the field), and the last one is the action of the field itself (a functional of the field only). In this case we know that all the classical electromagnetic theory can be deduced from the action (2.1) and we can also obtain an acceleration $\dot{u}^i = du^i/ds$ that satisfies

$$\eta_{ij} u^i \dot{u}^j = 0 \quad \text{or} \quad \frac{d}{ds}(\eta_{ij} u^i u^j) = 0 \qquad (2.2)$$

a condition that must necessarily be fulfilled in flat space-time because from

$$ds = (\eta_{ij} dx^i dx^j)^{1/2} \qquad (2.3)$$

we can deduce equation (2.2).

Let us now consider the gravitational case as generally as we can. So let us state that space-time is only a differential manifold, with generic coordinates $x^i$. We shall use an arbitrary system of coordinates, so all our equations must be covariant. Let $s$ be a general parameter and $u^i = dx^i/ds$. As in the electromagnetic case we take a gravitational action:

$$S_g = \sum \int m g_{ij}(x^i) u^i u^j ds + \sum \int m \mathcal{L}^{(I)}(x^i, u^i) ds + \int \mathcal{L}^{(F)} d^4 x \qquad (2.4)$$

where the role played by the three terms is the same as in equation (2.1), $m$ is the mass particle, and $g_{ij}(x)$ are only arbitrary coefficients. Since we want our theory to be as general as possible we make the least possible number of hypotheses concerning the interaction Lagrangian $\mathcal{L}^{(I)}$ which is a function of $u^i$ and $x^i$ (through some unspecified tensor and spinor fields, so our theory has, so far, an arbitrary spin). We only ask that $\mathcal{L}^{(I)}$ not be an explicit function of $s$, and this is its sole symmetry.

To obtain the equation of motion of a particle we make a virtual displacement $\delta$ of the space-time path of the particle, leaving the eventual field fixed. Then the one-particle equation of motion must be such that

$$\delta S_g = m \delta \int g_{ij}(x^i) u^i u^j ds + m \delta \int \mathcal{L}^{(I)} ds = 0 \qquad (2.5)$$

We have placed the same mass $m$ in the first and second term of the right-hand side of the equation (2.4) so both cancel out in equation (2.5). In this way the weak equivalence principle is satisfied by the equation of motion, which turns out to be $m$ independent.

From now on we shall work with the Lagrangian

$$\mathcal{L} = \mathcal{L}^{(P+I)} = g_{ij} u^i u^j + \mathcal{L}^{(I)} \qquad (2.6)$$

Then we can use the Euler–Lagrange equation to write the motion equations of motion explicitly:

$$\frac{d}{ds}\left(\frac{\partial \mathcal{L}}{\partial u^i}\right) - \frac{\partial \mathcal{L}}{\partial x^i} = 0 \qquad (2.7)$$

If we multiply this equation by $u^i$, and we follow the well-known steps of classical mechanics, we have

$$\frac{dH}{ds} = 0, \quad \text{where } H = \frac{\partial \mathcal{L}}{\partial u^i} u^i - \mathcal{L} \qquad (2.8)$$

so the function $H$ is a constant of the motion along the space-time path of the particle. Moreover it is the only constant of the motion because the only symmetry of $\mathcal{L}$ is that it is not an explicit function of $s$, and we know by Noether's theorem that each symmetry produces a constant of the motion and vice versa.

Therefore $H$ is the constant of the motion which follows from $\mathcal{L}$'s only symmetry.

### §(3): First Theorem: Flat Space-Time Conditions (cf. [5])

If we want to work in a flat space-time, we must choose an $\mathcal{L}$ that leads to a flat space-time without the use of any new hypothesis, i.e., a system of coordinates $x^i$ must exist, in our differentiable manifold, such that a constant metric tensor $\eta_{ij}$ may exist, and such that the world interval is

$$ds^2 = \eta_{ij} dx^i dx^j \quad \text{or} \quad \eta_{ij} u^i u^j = 0 \tag{3.1}$$

and of course

$$\frac{d}{ds}(\eta_{ij} u^i u^j) = 0 \tag{3.2}$$

*Theorem 1.* A constant metric tensor $\eta_{ij}$ with the required properties exists if and only $\mathcal{L}$ has the form

$$\mathcal{L} = \mathcal{L}^{(1)}(x^i, u^i) + F(\eta_{ij} u^i u^j) \tag{3.3}$$

where $\mathcal{L}^{(1)}$ is an homogeneous function of degree 1, in the variables $u^i$, $F$ is an arbitrary function but $\neq \text{const}\,(\eta_{ij} u^i u^j) + \text{const}$, and $\eta_{ij}$ is a constant tensor in the coordinates $x^i$.

*Proof.* In fact, the condition is sufficient because equations (2.8) and (3.3) lead to

$$x^{3/2} \frac{d}{dx}\left[\frac{F(x)}{x^{1/2}}\right] = H = \text{const} \tag{3.4}$$

where $x = \eta_{ij} u^i u^j$ and $F(x) \neq \text{const}\, x^{1/2} + \text{const}$, so we have that

$$x = \eta_{ij} \frac{dx^i}{ds} \frac{dx^i}{ds} = \text{const} \tag{3.5}$$

Then $x = \text{const}$ along every particle's space-time path, i.e., every geodesic. Therefore if we make $x = 1$ at a point $x_0^i$ for every $dx^i/ds$ (i.e., we fix a measure unit at $x_0^i$) and afterwards we link $x_0^i$ to a generic point $x^i$ with a geodesic we shall have $x = 1$ at $x^i$ and so $ds^2 = \eta_{ij} dx^i dx^j$.

Condition (3.3) is also necessary because if space-time is flat $x = \eta_{ij} u^i u^j$ is a constant of the motion. But, for a generic Lagrangian $\mathcal{L}$, with the only symmetry that it is not an explicit function of $s$, the only constants of the motion are $H$ and obviously all functions of $H$, so $x$ must be a function of $H$, or inversely

$$H = \frac{\partial \mathcal{L}}{\partial u^i} u^i - \mathcal{L} = f(x) \tag{3.6}$$

where $f$ is an arbitrary function of $x$, but of course $f(x) \neq$ const. But equation (3.6) is an inhomogeneous linear differential equation whose solution is

$$\mathcal{L} = \mathcal{L}^{(1)} + \frac{x^{1/2}}{2} \int \frac{f(x)}{x^{3/2}} dx \tag{3.7}$$

where $\mathcal{L}^{(1)}$ is the general solution of the homogeneous differential equation (3.6) and therefore an arbitrary but homogeneous function of the $u^i$ of degree 1, and the second term is a particular solution of equation (3.6). As $f(x) \neq$ const this second term is different from "const $x^{1/2}$ + const." □

Of course, in this case $\mathcal{L}^{(1)}(x^i, u^i)$ must also be a Lorentz scalar so it must be a function of the coordinates $x^i$ through a set of tensor and spinor fields.

The first example of a Lagrangian like (3.3) is the electromagnetic Lagrangian. On the other hand the only gravitational theory that has a Lagrangian that is similar to (3.3) is that of Belinfante (cf. [6]). Mavrides [7] considers this attempt as the only feasible ("viable") flat space-time theory.

### §(4):   Second Theorem: The Strong Equivalence Principle Condition

The strong equivalence principle states that in every point $x_0^i$ of space-time a system of coordinates $x^{i'}$ must exist where locally gravitational forces vanish, so at that point $x_0^{i'}$ the space-time paths of particles must behave locally as straight lines, for all directions $u^{i'}$, i.e., if $s$ is an arbitrary parameter,

$$\dot{u}^{i'} = \frac{du^{i'}}{ds} = \lambda u^{i'}, \quad \text{where } u^{i'} = \frac{dx^{i'}}{ds} \tag{4.1}$$

with an adequate parameter we shall have,

$$\dot{u}^{i'} = 0 \tag{4.2}$$

If we write equation (4.1) in an arbitrary system of coordinates $x^i$, making a change of coordinates

$$x^i = x^i(x^{i'}) \tag{4.3}$$

we shall have

$$\dot{u}^i + \Gamma_{kl}^i(x_0^i) u^k u^l = 0 \tag{4.4}$$

where $\Gamma_{kl}^i(x_0^i)$ are coefficients that depend on the system of coordinates, and they are, in fact, the coefficients of an affine symmetric connection because they behave as such under a change of coordinates. Therefore the strong equivalence principle defines a symmetric affine connection over our differential manifold. The $\Gamma_{kl}^i$ are point functions through some unspecified field and its derivatives.

But (4.4) is the equation of motion of the particle. Therefore if we want our theory to fulfill the strong equivalence principle we must choose $\mathcal{L}$ in such a way that the Euler-Lagrange equations (2.7) and (4.4) should be equivalent.

We shall prove Theorem 2 only within the set of Lagrangians of the form

$$\mathcal{L} = \sum_{-\infty}^{\infty} \mathcal{L}^{(n)} \tag{4.5}$$

where each term of the right-hand side is a homogeneous function of degree $n$ in the variables $u^i$, i.e.,

$$\frac{\partial \mathcal{L}^{(n)}}{\partial u^i} u^i = n \mathcal{L}^{(n)} \tag{4.6}$$

Notwithstanding, this set is large enough to contain all the Lagrangians used until now in gravitational theories. We shall generalize the theorem elsewhere.

As we wish the principle of general relativity to be satisfied, the equation of motion should be covariant. So $\mathcal{L}$, and also all the $\mathcal{L}^{(n)}$, must be scalar.

Under these conditions we can state the following theorem.

*Theorem 2.* $\mathcal{L}$ satisfies the strong equivalence principle if and only if

$$\mathcal{L} = g_{ik}(x^i) u^i u^k + \frac{\partial G(x^i)}{\partial x^i} u^i \tag{4.7}$$

where $g_{ik}(x^i)$ is a point function tensor and $G(x^i)$ is an arbitrary point function scalar.[1]

*Proof.* In fact: As $\mathcal{L}$ is not an explicit function of $s$ we can write the Euler–Lagrange equations (2.7) as

$$\frac{\partial^2 \mathcal{L}}{\partial u^k \partial u^i} \dot{u}^k + \frac{\partial^2 \mathcal{L}}{\partial x^k \partial u^i} u^k - \frac{\partial \mathcal{L}}{\partial x^i} = 0 \tag{4.8}$$

Then if we replace (4.5) and (4.6) in (4.8) we obtain

$$\frac{\partial^2 \mathcal{L}}{\partial u^k \partial u^i} \dot{u}^k + \left( \frac{\partial^2 \mathcal{L}^{(1)}}{\partial x^k \partial u^i} u^k - \frac{\partial \mathcal{L}^{(1)}}{\partial x^i} - \frac{\partial \mathcal{L}^{(0)}}{\partial x^i} \right) + B'_{lki} u^l u^k = 0 \tag{4.9}$$

where

$$B'_{lki} = B'_{kli} = \sum_{n \neq 1} \frac{1}{(n-1)} \frac{\partial^3 \mathcal{L}^{(n)}}{\partial x^{(k} \partial u^{l)} \partial u^i} - \sum_{n \neq 0,1} \frac{1}{n(n-1)} \frac{\partial^3 \mathcal{L}^{(n)}}{\partial x^i \partial u^k \partial u^l} \tag{4.10}$$

is a symmetric coefficient in $(l, k)$. We also want to write the three terms in large parentheses in equation (4.9) as a geometric object multiplied by $u^l u^k$, although in this case we cannot use equation (4.6); therefore we define an homogeneous

---

[1] $G(x^i)$ is, of course irrelevant for the variational principle.

function of degree 2 in the variable $u^i$, so that we can write

$$\mathcal{L}^{(0)} = \frac{\mathcal{L}^{(0)}}{L^{(2)}} L^{(2)} = L^{(-2)} L^{(2)} \tag{4.11}$$

$$\mathcal{L}^{(1)} = \frac{\mathcal{L}^{(1)}}{L^{(2)}} L^{(2)} = L^{(-1)} L^{(2)} \tag{4.12}$$

Now it is possible to use equation (4.6) on $L^{(-1)}$ and $L^{(-2)}$ (because $-1, -2 \neq 0, 1$), and we obtain

$$-\frac{\partial \mathcal{L}^{(0)}}{\partial x^i} = B''_{lki} u^l u^k \tag{4.13}$$

where

$$B''_{lki} = \frac{1}{6} \frac{\partial}{\partial x^i} \left( L^{(2)} \frac{\partial^2 L^{(-2)}}{\partial u^l \partial u^k} \right) \tag{4.14}$$

and

$$\frac{\partial^2 \mathcal{L}^{(1)}}{\partial x^k \partial u^i} u^k - \frac{\partial \mathcal{L}^{(1)}}{\partial x^i} = B'''_{lki} u^l u^k \tag{4.15}$$

where

$$B'''_{lki} = \frac{\partial}{\partial x^{(k}} \left( -\frac{L^{(2)}}{2} \frac{\partial^2 L^{(-1)}}{\partial u^{l)} \partial u^i} + L^{(-1)} \frac{\partial^2 L^{(2)}}{\partial u^{l)} \partial u^i} \right) - \frac{1}{2} \frac{\partial}{\partial x^i} \left( L^{(2)} \frac{\partial^2 L^{(-1)}}{\partial u^l \partial u^k} \right) \tag{4.16}$$

If we place equations (4.13)-(4.16) in equation (4.9) we have

$$\frac{\partial^2 \mathcal{L}}{\partial u^k \partial u^i} \dot{u}^k + B_{lki} u^l u^k = 0 \tag{4.17}$$

where

$$B_{lki} = B'_{lki} + B''_{lki} + B'''_{lki} \tag{4.18}$$

In the general case we can find the symmetric matrix $A^{il}$ such that

$$A^{il} \frac{\partial^2 \mathcal{L}}{\partial u^l \partial u^k} = \delta^l_k \tag{4.19}$$

Thus equation (4.17) becomes

$$\dot{u}^l + A^{li} B_{ski} u^s u^k = 0 \tag{4.20}$$

This is only the Euler-Lagrange equations (2.7) in a covariant form. This equation must be equivalent to the equations of motion (4.4) in order to fulfill the strong equivalence principle; therefore

$$A^{li} B_{ski} = \Gamma^l_{sk} \tag{4.21}$$

But $A^{li}$ and $B_{ski}$ are functions of $x^i$ and $u^i$ while $\Gamma^l_{sk}$ is only a function of $x^i$, therefore the strong equivalence principle is fulfilled if and only if

$$\frac{\partial A^{li}}{\partial u^m} B_{ski} + A^{li} \frac{\partial B_{ski}}{\partial u^m} = 0 \quad \forall u^i \tag{4.22}$$

Now, from equations (4.19) and (4.22) we conclude that necessarily

$$\frac{\partial B_{skr}}{\partial u^m} - \frac{\partial B_{skm}}{\partial u^r} = 0 \tag{4.23}$$

From now on we shall work only with this necessary condition.

Replacing (4.18) in equation (4.23) and equating to zero[2] all terms having a different degree in $u^i$, we obtain

$$\frac{\partial^4 \mathcal{L}^{(n)}}{\partial u^r \partial x^m \partial u^k \partial u^s} - \frac{\partial^4 \mathcal{L}^{(n)}}{\partial u^m \partial x^r \partial u^k \partial u^s} = 0, \quad n \neq 0, 1 \tag{4.24}$$

$$\frac{\partial B''_{skr}}{\partial u^m} - \frac{\partial B''_{skm}}{\partial u^r} = 0, \quad n = 0 \tag{4.25}$$

$$\frac{\partial B'''_{skr}}{\partial u^m} - \frac{\partial B'''_{skm}}{\partial u^r} = 0, \quad n = 1 \tag{4.26}$$

we can integrate equation (4.24) twice in $u^i$ and obtain

$$\frac{\partial^2 \mathcal{L}^{(n)}}{\partial x^m \partial u^r} - \frac{\partial^2 \mathcal{L}^{(n)}}{\partial x^r \partial u^m} = D_{rmk}(x^i) u^k + F_{rm}(x^i) \tag{4.27}$$

where $D_{rmk}(x^i)$ and $F_{rm}(x^i)$ are two arbitrary antisymmetric point functions in the indices $(r, m)$. But as the left-hand side has a well-defined degree $(n-1)$ in $u^i$, the right-hand side must have the same degree, so we have three possibilities:

$$\begin{align} \text{(a)} \quad & D_{rmk} = 0; \quad F_{rm} \neq 0 \\ \text{(b)} \quad & D_{rmk} \neq 0; \quad F_{rm} = 0 \\ \text{(c)} \quad & D_{rmk} = 0; \quad F_{rm} = 0 \end{align} \tag{4.28}$$

(a) In this case equation (4.27) becomes a linear nonhomogeneous differential equation with the particular solution $\mathcal{L}^{(1)} = A_i(x^i) u^i$, but since equation (4.27) comes from equation (4.24), which is not valid for $n = 1$, this solution must be rejected. Then, we have, of course, the general solution of the corre-sponding homogeneous equations but we shall study this solution in case (c). So we have finished with case (a).

---

[2] If $\sum_n f^{(n)} = 0$, where $f^{(n)}(x)$ is a homogeneous function of degree $n$, we have that $f^{(n)}(\lambda x) = \lambda^n f^{(n)}_{(x)}$, therefore $\sum_n \lambda^n f^{(n)}(x) = 0$ for any $\lambda$ so $f^{(n)}(x) = 0$ for each $n$.

(b) In this case $F_{rm} = 0$, and so equation (4.27) becomes a nonhomogeneous linear differential equation with solution

$$\mathcal{L}^{(n)} = \mathcal{L}_p^{(n)} + \mathcal{L}_h^{(n)} \qquad (4.29)$$

where $\mathcal{L}_p^{(n)}$ is a particular solution and $\mathcal{L}_h^{(n)}$ is the general solution of the homogeneous equation. But we have only a nontrivial $\mathcal{L}_p^{(n)}$ if $n = 2$. Then

$$\mathcal{L}^{(n)} = \mathcal{L}^{(2)} = g_{ik}(x^i) u^i u^k + \mathcal{L}_h^{(2)} \qquad (4.30)$$

where $g_{ik}(x^i)$ is an arbitrary point function and $\mathcal{L}_h^{(2)}$ is an homogeneous function of degree 2 which is the solution of the homogeneous part of equation (4.27).

The homogeneous part of equation (4.27) has a first integral, which is

$$\frac{\partial \mathcal{L}_h^{(2)}}{\partial u^r} = \frac{\partial}{\partial x^r} (H + x^l F_l) = \frac{\partial G}{\partial x^r} \qquad (4.31)$$

where $H(x^i, u^i)$ and $F_l(u^i)$ are arbitrary functions. Therefore, if we multiply by $u^r$ and use (4.6) we have

$$\mathcal{L}_h^{(2)} = \frac{1}{2} \frac{\partial G}{\partial x^r} u^r \qquad (4.32)$$

Now, if we study equation (4.31) under a change of coordinates we see that its left-hand side is a vector and its right-hand side is a vector only if $G$ is a point function, but in this case the left-hand side has degree zero, so the only possible solution is

$$\mathcal{L}_h^{(2)} = 0 \qquad (4.33)$$

Therefore the most general solution in case (b) is

$$\mathcal{L}^{(2)} = g_{ik} u^i u^k \qquad (4.34)$$

Regarding case (c) we must solve the homogeneous part of equation (4.27). Following the same procedure, we reach the conclusion that the only possible scalar solution is

$$\mathcal{L}^{(1)} = \frac{\partial G(x^i)}{\partial x^r} u^r = \frac{dG(x^i)}{ds} \qquad (4.35)$$

but this solution must be rejected because equation (4.27) is not valid in the case $n = 1$, and anyhow it is only a total derivative with no effect on the equation of motion.

At this stage we have

$$\mathcal{L} = \mathcal{L}^{(0)} + \mathcal{L}^{(1)} + g_{ik}(x^i) u^i u^k \qquad (4.36)$$

where $\mathcal{L}^{(0)}$ and $\mathcal{L}^{(1)}$ must satisfy equations (4.25) and (4.26), respectively.

If we multiply equation (4.25) by $u^s u^k$ and use equations (4.13) and (4.14) we obtain

$$\frac{\partial}{\partial x^i}\left(\frac{\partial \mathcal{L}^{(0)}}{\partial u^s} + L^{(2)} \frac{\partial L^{(-2)}}{\partial u^s}\right) - \frac{\partial}{\partial x^s}\left(\frac{\partial \mathcal{L}^{(0)}}{\partial u^i} + L^{(2)} \frac{\partial L^{(-2)}}{\partial u^i}\right) \qquad (4.37)$$

whose integral is

$$\frac{\partial \mathcal{L}^{(0)}}{\partial u^i} + L^{(2)} \frac{\partial L^{(-2)}}{\partial u^i} = \frac{\partial G}{\partial x^i} \qquad (4.38)$$

where $G$ is similar to the function $G$ defined in equation (4.31). As $G$ is an arbitrary function we can choose a scalar one, therefore the left-hand side of equation (4.38) is a vector, and the right-hand side will be a vector only if $G$ is a point function. Then the left-hand side has degree 1 in the variables $u^i$ while the right-hand side has degree zero unless the left-hand side is zero, then if we multiply it by $u^i$ and use equation (4.6) we have that $\mathcal{L}^{(0)} = 0$. So we can eliminate the first term in the right-hand side of equation (4.36).

If we follow the same procedure with equation (4.26) after a long but elementary computation we have

$$\frac{\partial}{\partial x^s}\left(2\frac{\partial \mathcal{L}^{(1)}}{\partial u^i} - L^{(-1)}\frac{\partial L^{(2)}}{\partial u^i} + L^{(2)}\frac{\partial L^{(-1)}}{\partial u^i}\right)$$
$$- \frac{\partial}{\partial x^i}\left(2\frac{\partial \mathcal{L}^{(1)}}{\partial u^s} - L^{(-1)} \frac{\partial L^{(2)}}{\partial u^s} + L^{(2)} \frac{\partial L^{(-1)}}{\partial u^s}\right) = 0 \qquad (4.39)$$

whose integral is

$$2\frac{\partial \mathcal{L}^{(1)}}{\partial u^i} - L^{(-1)} \frac{\partial L^{(2)}}{\partial u^i} + L^{(2)} \frac{\partial L^{(-1)}}{\partial u^i} = \frac{\partial G}{\partial x^i} \qquad (4.40)$$

where $G$ is a function as in equation (4.31) and following the same steps we can see that $G$ must be a point function. Therefore multiplying (4.40) by $u^i$ we have

$$\mathcal{L}^{(1)} = -\frac{\partial G}{\partial x^i} u^i = -\frac{dG}{ds} \qquad (4.41)$$

so $\mathcal{L}$ must necessarily have the form (4.7) in order to fulfill the strong equivalence principle.

But obviously it is sufficient to give the Lagrangian $\mathcal{L}$ the form (4.7) to fulfill the strong equivalence principle. □

## §(5): Conclusion

As it is impossible to find a Lagrangian with both forms (3.3) and (4.7) we reach the conclusion stated in the Introduction: We cannot make a flat space-time gravitational theory based on a unique variational principle and fulfill the strong equivalence principle. Therefore in flat space-time we must follow one of these two ways:

1. Make a theory with no unique variational principle; but this theory would be even more exceptional than general relativity.
2. Make a theory with no strong equivalence principle.

These results, like every negative result, must be taken with great caution. But we think that if one postulates the strong equivalence principle and the relativity principle through a variational mechanism, curved space-time and Einstein's general relativity follow inevitably.